\documentclass[aps,prd,preprint,groupedaddress]{revtex4}
\usepackage{epsfig}

\usepackage{color}

\def\bea{\begin{eqnarray}}
\def\eea{\end{eqnarray}}

\def\pbp{\langle\bar{\psi}\psi\rangle}
\def\gsim{\mathrel{\rlap{\lower4pt\hbox{\hskip1pt$\sim$}}\raise1pt\hbox{$>$}}}

\begin{document}

\title{QCD thermodynamics with continuum extrapolated Wilson fermions II.}

\author{Szabolcs Borsanyi}
\email{borsanyi@uni-wuppertal.de}
\affiliation{University of Wuppertal, Department of Physics, Wuppertal D-42097, Germany}

\author{Stephan Durr}
\email{duerr@uni-wuppertal.de}
\affiliation{University of Wuppertal, Department of Physics, Wuppertal D-42097, Germany}
\affiliation{Julich Supercomputing Center, Forschungszentrum Julich, Julich D-52425, Germany}

\author{Zoltan Fodor}
\email{fodor@bodri.elte.hu}
\affiliation{University of Wuppertal, Department of Physics, Wuppertal D-42097, Germany}
\affiliation{Julich Supercomputing Center, Forschungszentrum Julich, Julich D-52425, Germany}
\affiliation{Eotvos University, Institute for Theoretical Physics, Budapest 1117, Hungary}

\author{Christian Holbling}
\email{hch@uni-wuppertal.de}
\affiliation{University of Wuppertal, Department of Physics, Wuppertal D-42097, Germany}

\author{Sandor D. Katz}
\email{katz@bodri.elte.hu}
\affiliation{Eotvos University, Institute for Theoretical Physics, Budapest 1117, Hungary}
\affiliation{MTA-ELTE Lendulet Lattice Gauge Theory Research Group, 1117 Budapest, Hungary}

\author{Stefan Krieg}
\email{s.krieg@fz-juelich.de}
\affiliation{University of Wuppertal, Department of Physics, Wuppertal D-42097, Germany}
\affiliation{Julich Supercomputing Center, Forschungszentrum Julich, Julich D-52425, Germany}

\author{Daniel Nogradi}
\email{nogradi@bodri.elte.hu}
\affiliation{Eotvos University, Institute for Theoretical Physics, Budapest 1117, Hungary}
\affiliation{MTA-ELTE Lendulet Lattice Gauge Theory Research Group, 1117 Budapest, Hungary}

\author{Kalman K. Szabo}
\email{szaboka@general.elte.hu}
\affiliation{University of Wuppertal, Department of Physics, Wuppertal D-42097, Germany}
\affiliation{Julich Supercomputing Center, Forschungszentrum Julich, Julich D-52425, Germany}

\author{Balint C. Toth}
\email{tothbalint@szofi.elte.hu}
\affiliation{University of Wuppertal, Department of Physics, Wuppertal D-42097, Germany}

\author{Norbert Trombitas}
\email{trombitas@bodri.elte.hu}
\affiliation{Eotvos University, Institute for Theoretical Physics, Budapest 1117, Hungary}
\affiliation{MTA-ELTE Lendulet Lattice Gauge Theory Research Group, 1117 Budapest, Hungary}

\begin{abstract}
We continue our investigation of $2+1$ flavor QCD thermodynamics using
dynamical Wilson fermions in the fixed scale approach.  Two additional pion
masses, approximately 440 MeV and 285 MeV, are added to our previous work at
545 MeV. The simulations were performed at 3 or 4 lattice spacings at each pion
mass.  The renormalized chiral condensate, strange quark number susceptibility
and Polyakov loop is obtained as a function of the temperature and we observe a
decrease in the light chiral pseudo-critical temperature as the pion mass is
lowered while the pseudo-critical temperature associated with the strange quark
number susceptibility or the Polyakov loop is only mildly sensitive to the pion
mass. These findings are in agreement with previous continuum results obtained
in the staggered formulation.
\end{abstract}


\maketitle
\section{Introduction}
\label{introduction}

The quantitative description of the quark gluon plasma (QGP) is in the focus
of the heavy ion program at the accelerators RHIC (Brookhaven) and LHC (CERN).
At the large energy densities achieved in these experiments quarks are no
longer confined into detectable particles (hadrons) but form a nearly ideal
fluid
\cite{Kumar:2013cqa}. The QGP phase is separated from the hot gas of
hadrons by a cross-over~\cite{Aoki:2006we} 
at high enough collision energies. This transition leaves an imprint in the
abundance of various particle species that are created at the break-up of the
plasma
\cite{BraunMunzinger:2003zd,Rafelski:2004dp,Andronic:2005yp,Cleymans:2005xv}
and the transition temperature can be modelled as a function of the collision
energy or baryo-chemical potential \cite{BraunMunzinger:2003zz,Andronic:2008gu,Abelev:2008ab,Becattini:2012xb,Becattini:2014hla}.

Lattice simulations provide an excellent method to solve the underlying
quantum field theory, Quantum Chromodynamics (QCD) in equilibrium
\cite{Petreczky:2012rq}.  Lattice calculations are valid and feasible both in
the hadronic and in the quark gluon phase, allowing a first principles
description of the transition itself.  The appeal of lattices methods includes
that no approximation is involved, the complete path integral of the
discretized theory is calculated. The features of the continuum theory can then
be obtained through continuum extrapolation from sufficiently high resolutions.

Contrary to experiments, lattice QCD has the advantage to
access many possible theories with various quark masses \cite{Brown:1990ev}.
E.g. it has been shown that QCD with infinite quark masses
exhibits a 1st order transition between QGP and the confined
phase  \cite{Celik:1983wz}, which is clearly signalled by the Polyakov loop,
the exponentialized single quark free energy: the Polyakov loop is non-zero
only for deconfined quarks. The first order nature persists when quarks become
dynamical but heavy \cite{Hasenfratz:1983ce}.  For very light quarks, the order
of the transition depends on the number of light flavors. The transition is
dominated by the restoration of chiral symmetry, signalled by the
vanishing of the chiral condensate. For intermediate masses there is no
real transition and both the Polyakov loop and the chiral condensate are approximate, remnant order parameters. In this work we will
study these for several sets of quark masses, complemented by a measure of
the confinement of the strange quarks, the strange quark number susceptibility.

Results with physical quark masses are abundant in the staggered formulation.
The order of the transition is cross-over \cite{Aoki:2006we} with a chiral
transition temperature $T_c\sim 155~\mathrm{MeV}$
\cite{Aoki:2006br,Aoki:2009sc,Borsanyi:2010bp,Bazavov:2011nk}. The equation of
state has been calculated with high precision
\cite{Borsanyi:2010cj,Borsanyi:2013bia,Bazavov:2014pvz}, even at small but
non-vanishing quark densities \cite{Borsanyi:2012cr,Hegde:2014sta} and there
exist predictions for the freeze-out parameters
\cite{Bazavov:2012vg,Borsanyi:2013hza,Borsanyi:2014ewa} and 
fluctuations of various conserved charges \cite{Borsanyi:2011sw,Bazavov:2012jq}.

Yet it is not certain whether all systematics are controlled in
staggered simulations. The staggered fermion action describes four flavors,
which reduces to a single quark flavor  by rooting the fermion sector,
thus potentially giving up locality. This conceptually uncertain
step is completely avoided in the Wilson formulation, which we also use in the
present work. The theoretical soundness comes at a price. The Dirac spinor of
Wilson fermions have four components, while the staggered spinor is a single
complex field.  More importantly chiral symmetry is explicitly  broken at
any lattice spacing, it is restored only in the continuum limit. This leads to a
more complicated structure of divergences, and a demand for fine lattices. The
use of a heavier than natural pion mass can significantly reduce the costs of
an individual simulation.
Indeed, although there exists zero temperature studies with Wilson fermions in
the physical point \cite{Durr:2010vn}, the description of the QCD transition
with physical Wilson quarks is still missing.

There are also formulations which maintain a lattice version of chiral symmetry.
These are computationally even more challenging than Wilson fermions. For
thermodynamics results using overlap fermions see Ref.~\cite{Borsanyi:2012xf}.
In the domain wall formulations physical quark masses have been recently
reached, albeit not yet in continuum limit
\cite{Buchoff:2013nra,Bhattacharya:2014ara}.

The aim of thermodynamics studies with Wilson quarks goes beyond the obvious
long-term goal of reaching the physical point.  Several groups have already
studied the chiral scaling with two flavors of Wilson quarks
\cite{AliKhan:2000iz,Bornyakov:2009qh,Bornyakov:2011yb,Ejiri:2009hq,Brandt:2012jc,Brandt:2013mba},
and also in the twisted mass formulation \cite{Burger:2011zc, Burger:2014xga}. The transition
temperature with infinitely heavy strange quarks monotonically decreases when
the pion mass is lowered. Extrapolations to the physical pion mass give a value
around $\sim 170~\mathrm{MeV}$ \cite{Ejiri:2009hq,Bornyakov:2011yb,Burger:2011zc, Burger:2014xga}.

Most works with 2+1 flavors of Wilson quarks address the phenomenology
of QGP e.g. the equation of state \cite{Umeda:2012er} (using a pion mass of
$m_\pi\sim 550~\mathrm{MeV}$). For some applications anisotropic lattices were
introduced and the quark number susceptibilities \cite{Giudice:2013fza} and
transport coefficients \cite{Amato:2013naa,Aarts:2014nba} have been calculated
($m_\pi\sim 392~\mathrm{MeV}$).  In \cite{Borsanyi:2011kg, Borsanyi:2012uq} we
have started a study of $2+1$ flavor QCD thermodynamics using the Wilson
fermion formulation with a fixed pion mass of $m_\pi\sim 545~\mathrm{MeV}$. A
careful continuum extrapolation was performed and the results were found in
agreement with the continuum extrapolated staggered simulations with equal pion
mass. We have worked out the details of the renormalization procedure and in
this work we reapply these for two new sets of quark masses. In this work we
calculate three quantities of interest, the chiral condensate, strange quark
number susceptibility and the Polyakov loop for $440$ and $285$ MeV pions.  A
continuum extrapolation is performed for both masses.

Similarly to other $N_f=2$ studies we observe a monotonic shift in the
chiral transition temperature as the physical point is approached.
The pion mass dependence in the strange quark number susceptibility and the
Polyakov loop is significantly milder.

The picture can only be complete if the temperature scans are shown together
with data at the physical point. Such simulations with Wilson quarks are beyond
our resources for now. Thus we use the continuum extrapolations from our
staggered program to illustrate our expectations.
This also shows that decreasing the pion mass further towards the physical
point decreases the pseudo-critical temperature associated with the light
chiral condensate whereas it does not substantially effect the
pseudo-critical temperature associated with the strange quark number
susceptibility. 

The organization of the paper is as follows. In section \ref{simulationsetup}
we summarize the simulation setup, parameters and algorithms that were used. In
section \ref{observables} the measured observables are given and their
renormalization properties are discussed. In section \ref{results} we
present the results of our investigations and we finally conclude in section \ref{summary}.

\section{Simulation setup, line of constant physics} \label{simulationsetup}

The Symanzik tree level improved action \cite{Symanzik:1983dc, Luscher:1984xn}
is used in the gauge sector while in the fermionic sector the clover
\cite{Sheikholeslami:1985ij} action further improved by six steps of stout
smearing is adopted \cite{Morningstar:2003gk}. The clover coefficient is set to
its tree level value $c_{SW} = 1$ and the stout smearing parameter is chosen at
$\varrho = 0.11$. For more details see \cite{Capitani:2006ni, Durr:2008rw} or
\cite{Borsanyi:2012uq} where the simulation setup was identical to the current
work except for the values of the quark masses.

The light quarks $u$ and $d$ are assumed to be degenerate and a $2+1$ flavor
algorithm is used. The HMC algorithm \cite{Duane:1987de} is adopted for the
light quarks and the RHMC algorithm \cite{Clark:2006fx} for the strange quark.
Various algorithmic improvements are applied for speeding up the simulation:
the Sexton-Weingarten multiple time scale integration \cite{Sexton:1992nu}, the
Omelyan integration scheme \cite{Takaishi:2005tz} and even-odd preconditioning
\cite{DeGrand:1988vx}.

Finite temperature is introduced as a finite Euclidean temporal extent of the
lattice. If the lattice is isotropic, i.e.
the lattice spacing is identical in all directions, 
the temperature $T=1/aN_t$ is set by the number of time slices
$N_t$. Leaving the bare parameters unchanged one can thus vary the temperature
by simulating at different values of $N_t$. To facilitate a continuum extrapolation 
three or four sets of bare parameters have to be determined, each corresponding
to a different lattice spacing, but otherwise to the same physics (in terms of
a selection of mass ratios). This is called the fixed-scale approach
\cite{Umeda:2008bd}.

An alternative approach (mainly used in quenched and staggered simulations
as well as in studies with exact chiral symmetry) keeps the number of
time slices ($N_t$) constant in a temperature scan and the lattice spacing
is varied continuously through the bare parameters to tune the temperature.

In the absence of additive divergences in the bare parameters and assuming
the feasibility of the interpolation of various counterterms the latter
approach has a clear advantage: the simulation temperature can be selected
without restriction. The somewhat low temperature resolution of the fixed-scale
approach could be trivially improved by anisotropy, but also by redefining
the temporal boundary conditions \cite{Giusti:2010bb,Umeda:2014ula}.

In the Wilson formulation the additive divergences prevent the easy
interpolation of zero temperature data.  Repeated simulations at zero temperature are
costly, (even more so if the action is anisotropic \cite{Edwards:2008ja}) and
one uses as few sets of bare parameters as possible, leading to the fixed scale
approach, that we also use in this work in an isotropic setting.  In many cases
authors even forego the continuum extrapolation to spare the extra effort from
the use of several parameter sets.

\begin{table}
\begin{center}
\begin{tabular}{|c|c|c|c|c|}
\hline
$\beta$ & $a m_{ud}$ & $a m_s$  & $N_s$ & $N_t$ \\
\hline
\hline
$\;$3.30$\;$ &  $\;$-0.1122$\;$ &   $\;$-0.0710$\;$ & $\;$32$\;$ & $\;$6 - 16, $\;$32$\;$ \\
\hline                                                                               
$\;$3.57$\;$ &  $\;$-0.0347$\;$ &   $\;$-0.0115$\;$ & $\;$48$\;$ & $\;$6 - 16, $\;$64$\;$ \\
\hline                                                                               
$\;$3.70$\;$ &  $\;$-0.0181$\;$ &   $\;$0.0    $\;$ & $\;$48$\;$ & $\;$8 - 24, $\;$48$\;$ \\
\hline                                                                               
$\;$3.85$\;$ &  $\;$-0.0100$\;$ &   $\;$0.0050 $\;$ & $\;$64$\;$ & $\;$8 - 36, $\;$64$\;$ \\
\hline
\hline
$\beta$ & $a m_{ud}$ & $a m_s$  & $N_s$ & $N_t$ \\
\hline
\hline
3.30 &  -0.1245 &   -0.0710 & 32 & 6 - 16, 32 \\
\hline  
3.57 &  -0.0443 &   -0.0115 & 48 & 8 - 24, 64 \\
\hline  
3.70 &  -0.0258 &   0.0     & 64 & 8 - 24, 96 \\
\hline
\end{tabular}
\end{center}
\caption{Bare parameters for the $440$ MeV pion mass (top) and the $285$ MeV pion mass (bottom) simulations. 
The $N_t$ values used for the finite temperature runs and the values used
for the zero temperature runs are separated by a comma.}
\label{tab:parameters}
\end{table}

In this work we are working with four lattice spacings determined by
the inverse gauge coupling $\beta$. As in Ref.~\cite{Borsanyi:2012uq} we
use $\beta=3.30$, $3.57$, $3.70$ and $3.85$ corresponding
to lattice spacings from about $0.13~\mathrm{fm}$ to $0.05~\mathrm{fm}$.  The
scale was set by $m_{\Omega} = 1672~\mathrm{MeV}$. The temperature at each
fixed bare coupling $\beta$ is varied in discrete steps by varying $N_t$.

In our past work \cite{Borsanyi:2012uq} the pion mass was relatively heavy,
around $545~\mathrm{MeV}$.
Two sets of simulations were performed in the current work each corresponding 
to a fixed $m_\pi / m_\Omega$ and $m_K / m_\Omega$
mass ratio. In the first set the quark masses were tuned to 
$m_\pi / m_\Omega \simeq 0.26$ and $m_K / m_\Omega \simeq 0.34$. 
These correspond to about $m_\pi = 440$ MeV and $m_K = 570$ MeV. At this pion
mass the simulations were performed at all 4 lattice spacings. Finite volume
effects are expected to be small since $m_\pi L > 7$ at each lattice spacing.

In the second set the meson masses were tuned to $m_\pi / m_\Omega \simeq 0.17$
and $m_K / m_\Omega \simeq 0.32$, corresponding to about $m_\pi = 285$ MeV and
$m_K = 525$ MeV. At these pion masses the simulations were performed at 3
lattice spacings and for the finite volume of the system $m_\pi L > 5.4$ holds.

At each lattice spacing, i.e. fixed $\beta$, the mass of the strange quark
$m_s$ is fixed at its physical value across all three pion masses and the
physical point would be approached by changing $m_{ud}$ only. Hence as $m_{ud}$
is lowered, both $m_\pi$ and $m_K$ decrease towards their physical values.

\begin{figure}
\begin{center}
\includegraphics[width=8.5cm]{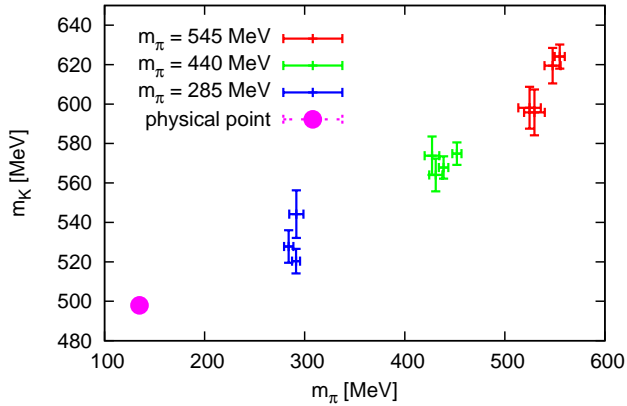} 
\caption{The various pion and kaon masses used in our past and current work. The heaviest pion mass is from 
our past work \cite{Borsanyi:2012uq}, the 4 red data points correspond to 4 lattice spacings. For the $m_\pi = 440$ MeV
point also 4 lattice spacings are used, while for the lightest pion mass, $m_\pi = 285$ MeV we have simulated at 3 lattice
spacings. The physical point is also shown for comparison. The scale is set by $m_{\Omega} = 1672$ MeV. \label{scatter}}
\end{center}
\end{figure}

A summary of the various pion and kaon masses used in our past and current work is shown in figure \ref{scatter}.
The bare quark masses, spatial and temporal lattice extents are shown in table \ref{tab:parameters} while
the measured meson, baryon and PCAC masses are shown in table \ref{tab:masses}. As can be seen $m_{\Omega}$ and hence the 
lattice spacing depends rather mildly on the light quark masses.
At each finite temperature point around 1000-1500 equilibrated unit length
trajectories were generated while we collected around 1000 trajectories at the zero temperature points. 
Autocorrelation times are around 5 - 25 trajectories close to the transition temperature depending on the
quantity, lattice spacing and pion mass.

\begin{table}
\begin{center}

\begin{tabular}{|c|c|c|c|c|c|}
\hline
$\beta$ & $m_\pi/m_\Omega$  & $m_K/m_\Omega$  & $a m_{PCAC}$   & $a m_\Omega$    & $a\; [fm]$ \\
\hline                                                                        
\hline                                                                        
$\;$3.30$\;$ &  $\;$0.262(3)$\;$  &    $\;$0.340(3)$\;$     &    $\;$0.0248(2) $\;$  & $\;$1.11(1) $\;$  & $\;$0.133(1)$\;$  \\
\hline                                                                                                              
$\;$3.57$\;$ &  $\;$0.270(3)$\;$  &    $\;$0.344(3)$\;$     &    $\;$0.01710(5)$\;$  & $\;$0.737(7)$\;$  & $\;$0.088(1)$\;$ \\
\hline                                                                                                              
$\;$3.70$\;$ &  $\;$0.258(4)$\;$  &    $\;$0.337(5)$\;$     &    $\;$0.01266(3)$\;$  & $\;$0.578(8)$\;$  & $\;$0.069(1)$\;$  \\
\hline                                                                                                              
$\;$3.85$\;$ &  $\;$0.256(4)$\;$  &    $\;$0.343(6)$\;$     &    $\;$0.00890(1)$\;$  & $\;$0.446(7)$\;$  & $\;$0.053(1)$\;$ \\
\hline
\hline                                                                        
$\beta$ & $m_\pi/m_\Omega$  & $m_K/m_\Omega$  & $a m_{PCAC}$   & $a m_\Omega$    & $a\; [fm]$ \\
\hline                                                                        
\hline                                                                        
3.30 &     0.174(4)          &   0.325(7)      &   0.0084(2)  & 0.97(2)     &  0.117(3) \\
\hline                                                                             
3.57 &     0.174(2)          &   0.311(4)      &   0.00693(4) & 0.723(8)    &  0.087(1)\\
\hline                                                                             
3.70 &     0.170(1)          &   0.316(5)      &   0.00481(2) & 0.560(9)    &  0.067(1)\\
\hline
\end{tabular}

\end{center}
\caption{Spectroscopy and physical scale results from zero 
temperature simulations, top: $m_\pi = 440$ MeV, bottom: $m_\pi = 285$ MeV.
The lattice spacings are set by $m_\Omega = 1672$ MeV.}
\label{tab:masses}
\end{table}

\section{Observables}
\label{observables}

The temperature dependencies of three quantities are determined in the current work, the renormalized light chiral condensate,
the strange quark number susceptibility and the renormalized Polyakov loop. 

\subsection{Chiral condensate}

The bare light chiral condensate requires both
additive and multiplicative renormalization. The details of the full renormalization procedure 
are given in \cite{Borsanyi:2012uq} following the references \cite{Bochicchio:1985xa, Giusti:1998wy} and will be
summarized below.

Additive renormalization at $T>0$ is implemented by the subtraction of $T=0$ quantities as this difference is free from
polynomial divergences in the inverse of the lattice spacing. Multiplicative renormalization is then achieved 
by the multiplication of the PCAC mass
$m_{PCAC}$ and the finite renormalization constant $Z_A$. The latter were determined in the chiral limit from 3-flavor
simulations in \cite{Borsanyi:2012uq} along the lines of \cite{Durr:2010vn, Durr:2010aw}
and can be taken from there directly for each $\beta$. Finally the Ward identity
establishes a relationship \cite{Giusti:1998wy} between the chiral condensate and the integrated pion correlator
leading to the final expression for the fully renormalized condensate at finite temperature,
\bea
\label{finalfinal}
m_R \pbp_R(T) &=& 2 N_f m_{PCAC}^2 Z_A^2 \Delta_{PP}(T)\;, 
\eea
where,
\bea
\label{diffpp}
\Delta_{PP}(T) = \int d^4 x \langle P_0(x) P_0(0) \rangle(T) - \int d^4 x \langle P_0(x) P_0(0) \rangle(T=0)
\eea
where $P_0(x)$ is the bare pseudo-scalar density; for more details see \cite{Borsanyi:2012uq}. The final result in
\cite{Borsanyi:2012uq} was shown for $m_R \pbp_R(T) / m_\pi^4$ since this combination is dimensionless. However when
comparing different pion masses as in the current work this normalization is not convenient because it introduces an
artificial pion mass dependence through the $4^{th}$ power. It turns out that the normalization
$m_R \pbp_R(T) / m_\pi^2 / m_{\Omega}^2$ is more suitable. This is because according to the GMOR relation at $T=0$,
the quark mass times the chiral condensate is proportional to $m_\pi^2$ to lowest order in chiral perturbation theory.
All results
related to the chiral condensate will be presented with the latter normalization and also the final result in
\cite{Borsanyi:2012uq} will be converted into it for comparison.

\subsection{Strange quark number susceptibility}

The strange quark number susceptibility $\chi_s = T/V\, \partial^2 \log Z / \partial \mu_s^2$, where $\mu_s$ is the strange quark
chemical potential, can be made dimensionless by considering $\chi_s / T^2$ and can be improved at
tree level by the division of its infinite volume and massless Stefan-Boltzmann limit at each finite $N_t$. The Stefan-Boltzmann
values for each $N_t$ were listed in \cite{Borsanyi:2012uq}. Furthermore $\chi_s / T^2$ is a finite quantity in the
continuum hence does not require any further renormalization factors. The strange quark number susceptibility is sensitive to the
confinement-deconfinement temperature of the strange quark and as we will see is only mildly dependent on the pion mass. 

\begin{figure}
\begin{center}
\includegraphics[height=7cm]{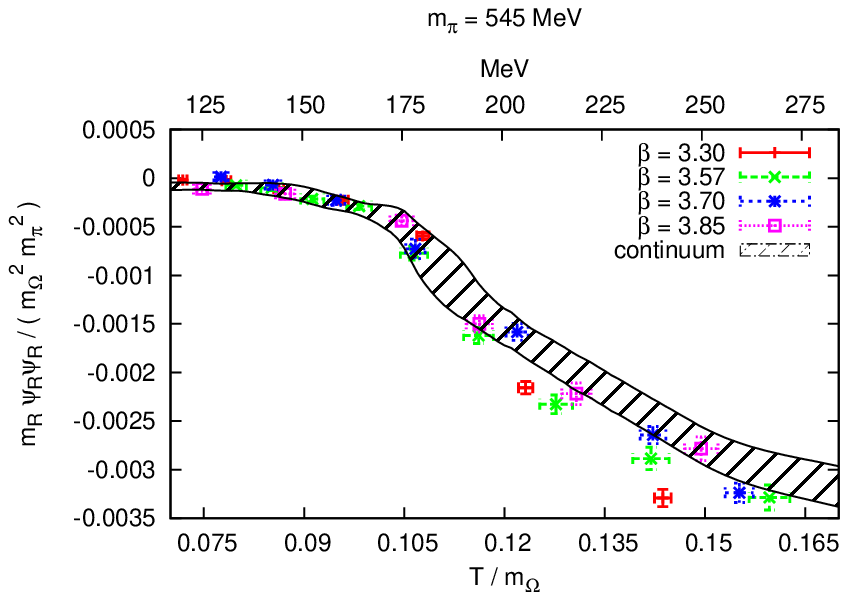}  

\includegraphics[height=7cm]{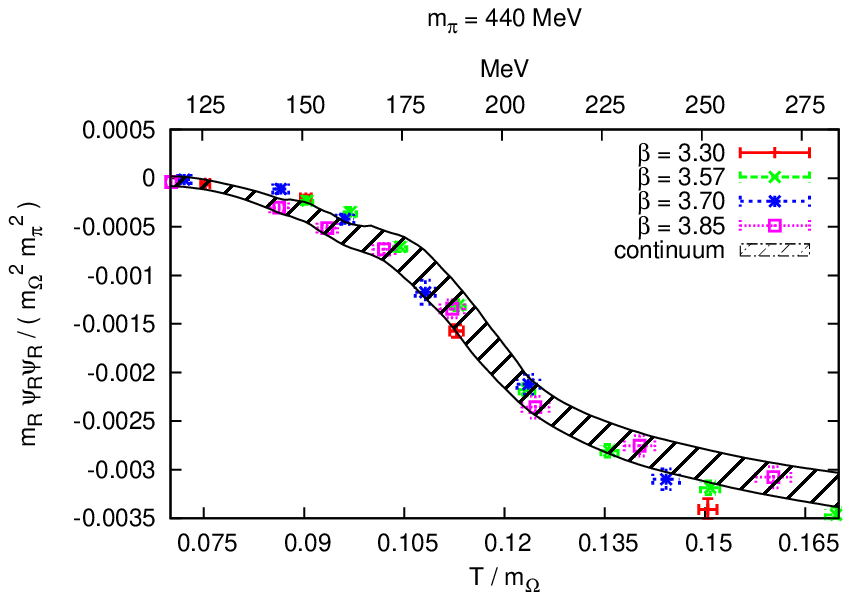} 

\includegraphics[height=7cm]{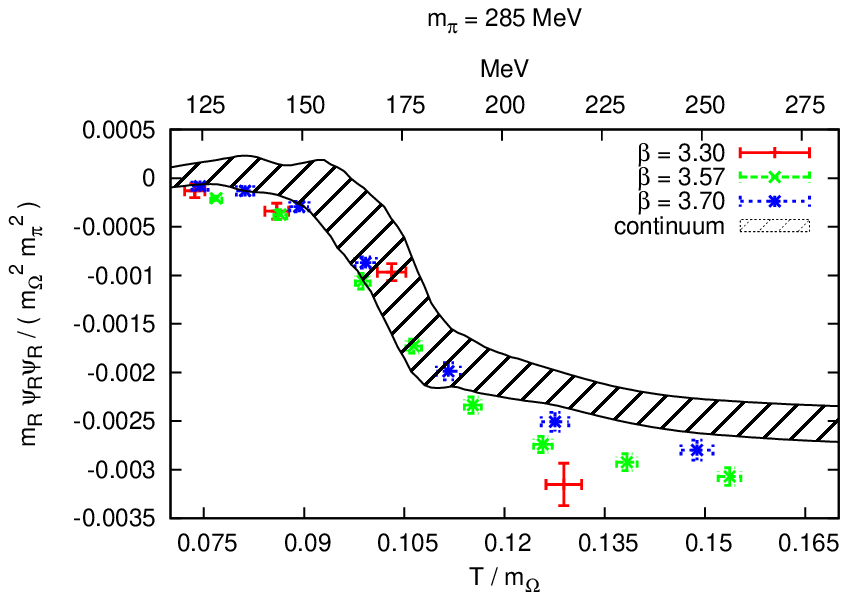} 
\caption{The renormalized chiral condensate for $m_\pi = 545$ MeV (top, from \cite{Borsanyi:2012uq}), 
$m_\pi = 440$ MeV (middle) and $m_\pi = 285$ MeV (bottom). The continuum
extrapolated results are also shown by the solid band.\label{pbp}}
\end{center}
\end{figure}

\subsection{Polyakov loop}

In order to renormalize the Polyakov loop one may use zero temperature quantities similarly to our description of the
renormalized chiral condensate \cite{Aoki:2006br}. However it is more convenient and 
less noisy to only use the finite temperature Polyakov
loop itself \cite{Gupta:2007ax}.

Our renormalization procedure for the Polyakov loop follows \cite{Borsanyi:2012uq}. The additive divergence 
of the free energy can be removed by the following renormalization prescription: a fixed value $L_*$ can be fixed for the
renormalized Polyakov loop at a fixed but arbitrary temperature $T_* > T_c$. This prescription leads to the following
renormalized Polyakov loop $L_R$ in terms of the bare quantity $L_0$,
\bea
\label{pr2}
L_R(T) = \left( \frac{L_*}{L_0(T_*)} \right)^{\frac{T_*}{T}} L_0(T)\;.
\eea
We choose $T_* = 0.143\; m_{\Omega}$ and $L_* = 1.2$ similarly to \cite{Borsanyi:2012uq} while other choices would simply
correspond to other renormalization schemes. For instance one may fix $T_*$ in units of $T_c$ as well.

\section{Results and continuum limit}
\label{results}

\begin{figure}
\begin{center}
\includegraphics[width=11cm]{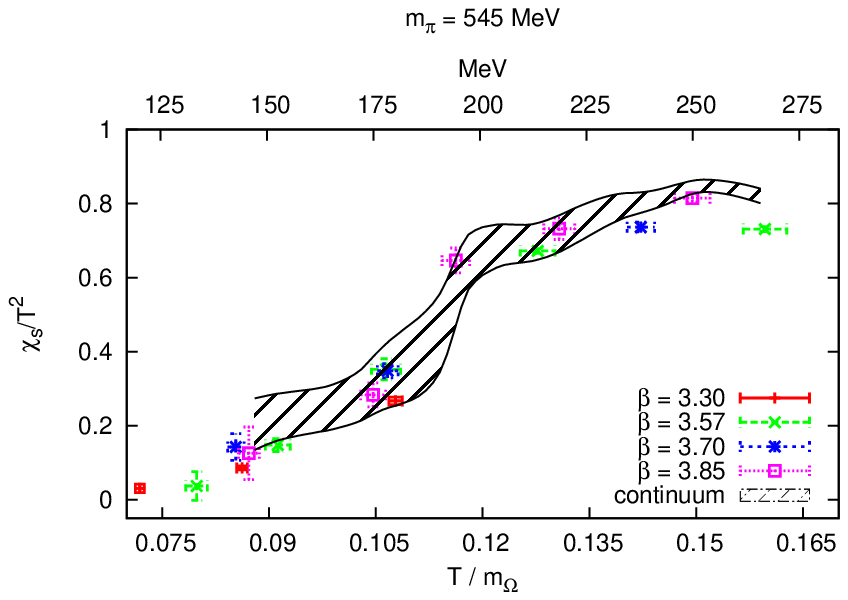}  

\includegraphics[width=11cm]{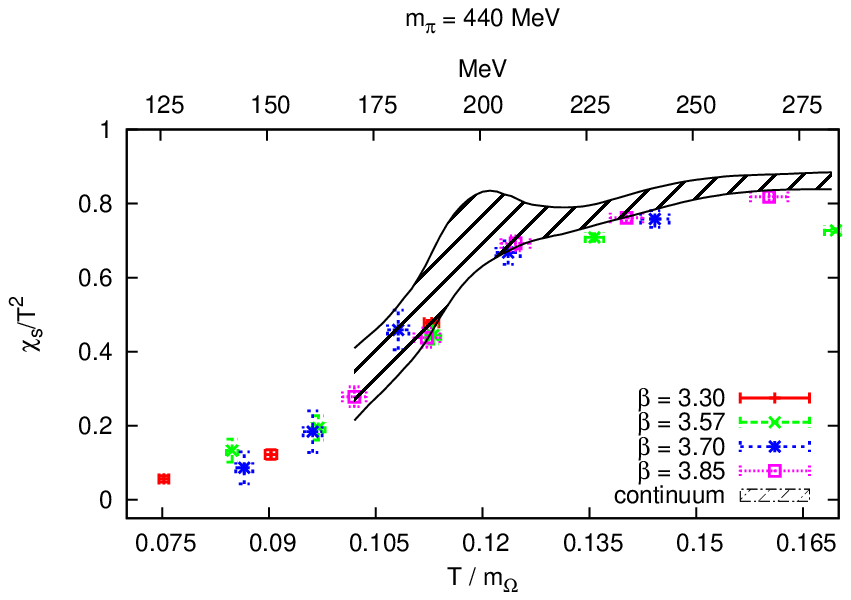}  

\includegraphics[width=11cm]{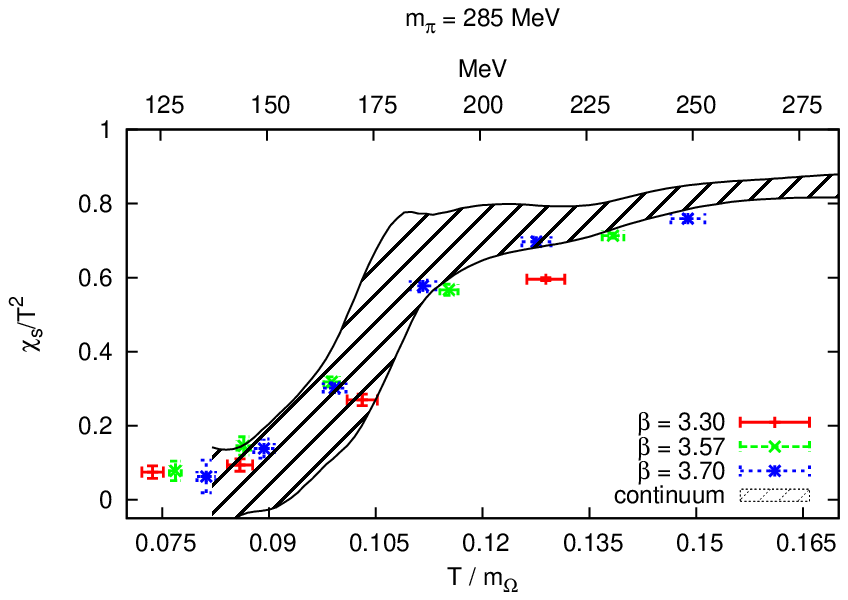} 
\caption{The strange quark number susceptibility for $m_\pi = 545$ MeV (top, from \cite{Borsanyi:2012uq}), 
$m_\pi = 440$ MeV (middle) and $m_\pi = 285$ MeV (bottom). The continuum
extrapolated results are also shown by the solid band.\label{qsusc}}
\end{center}
\end{figure}

At the $m_\pi = 440$ MeV point the simulations were performed at 4 lattice spacings while at the $m_\pi = 285$ MeV point
only at 3.
Since the fixed scale approach is used where the temperature can only be changed
by discrete amounts corresponding to the discrete changes in $N_t$ at each bare coupling $\beta$ an interpolation is
necessary in order to have a continuous curve as a function of temperature for each observable. In our previous work in
\cite{Borsanyi:2012uq} a spline interpolation was adopted and in the current work we add another method.

\begin{figure}
\begin{center}
\includegraphics[width=11cm]{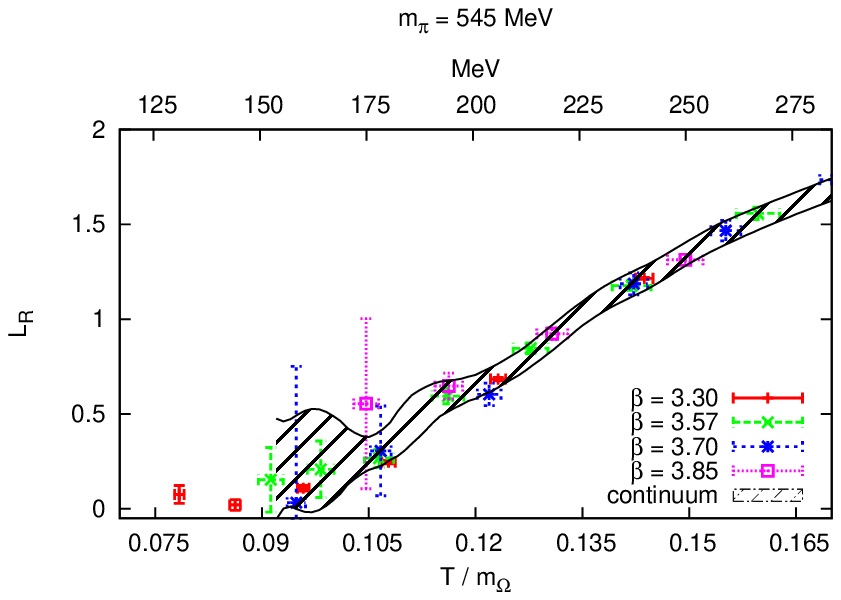}  

\includegraphics[width=11cm]{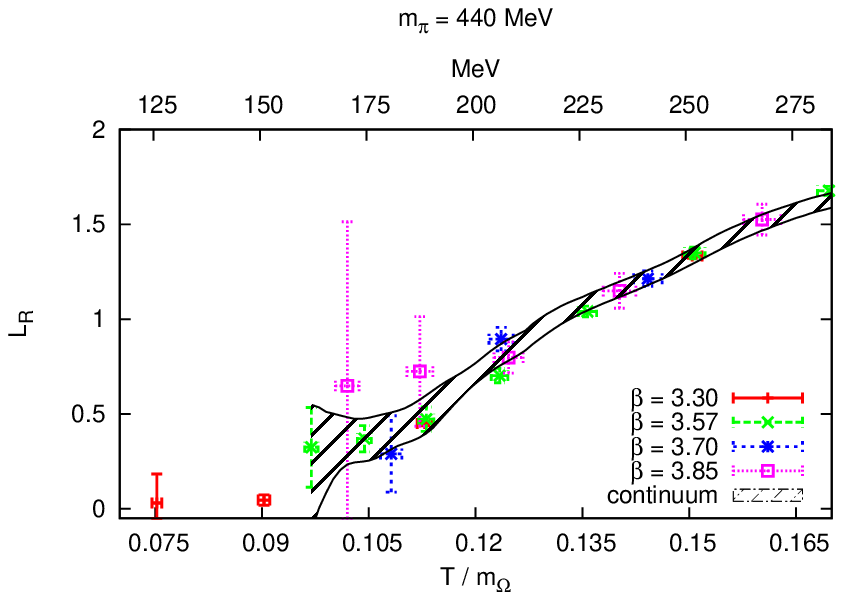}  

\includegraphics[width=11cm]{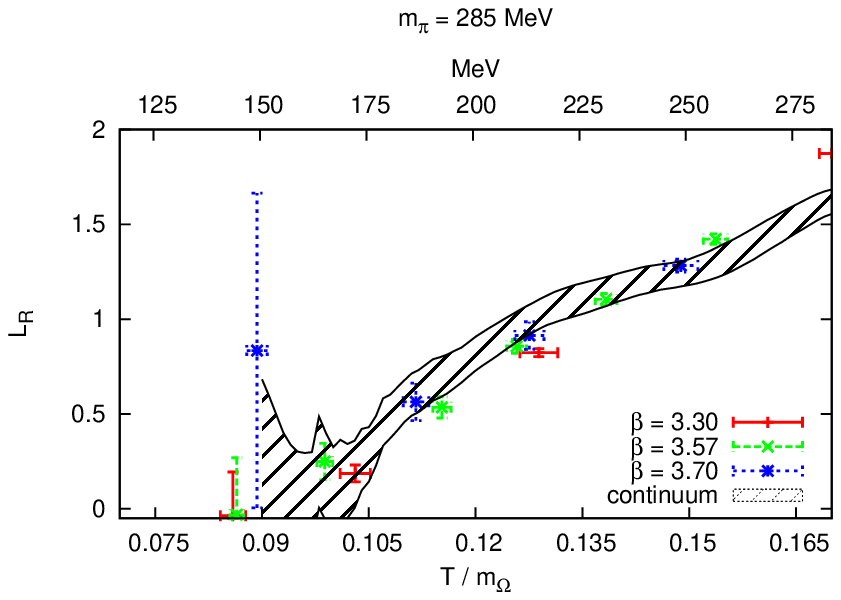} 
\caption{The renormalized Polyakov loop for $m_\pi = 545$ MeV (top, from \cite{Borsanyi:2012uq}), $m_\pi = 440$ MeV 
(middle) and $m_\pi = 285$ MeV (bottom) pion masses. The continuum
extrapolated results are also shown by the solid band.\label{ploop}}
\end{center}
\end{figure}

The previous method of \cite{Borsanyi:2012uq} consists of randomly placing node points for cubic spline interpolations
and the parameters of the spline are continuum extrapolated assuming $O(a)$ and $O(a^2)$ cut-off effects. Each result
corresponding to a fixed set of node points is weighted by its fit quality. The deviation
of the two continuum extrapolations as well as the spread with the various random choices for the 
nodes allow us to estimate systematic effects coming from both the continuum extrapolation and the interpolation. For
more details see \cite{Borsanyi:2012uq}.

In the current work we analyzed all of our data using a second method as well.
Clearly the expectation is that all quantities are monotonous functions of the temperature. This constraint is imposed
on our cubic spline fit following the algorithm \cite{MR567271}.
Naturally the statistical uncertainty of our measured data points will lead to a statistical uncertainty for the
interpolated curve. The obtained continuous interpolated curves and their error for each $\beta$ can then be used for a
continuum extrapolation at each $T$ (note that in the previous method the interpolation and extrapolation was performed
simultaneously in a global fit). We have performed the continuum extrapolation assuming both $O(a)$ an $O(a^2)$
cut-off effects but the $\chi^2/dof$ values of the $O(a^2)$ were much better hence completely dominate the final result
after performing an AIC weighted averaging \cite{aic3, aic4, aic1, aic2}.
The data from \cite{Borsanyi:2012uq} is reanalized in this slightly different way in
order to have a consistent analysis for all 3 pion masses and it
is reassuring to see that the final continuum results agree with the previous analysis in \cite{Borsanyi:2012uq}.

In order to assess the size of systematic uncertainties we perform all continuum fits by keeping all 4 lattice
spacings for the 2 heavier pion masses and also by dropping the roughest one, $\beta = 3.30$ and hence using only 3. 
The two continuum
results are then weighted according to \cite{aic3, aic4, aic1, aic2}. 
The deviation between the two continuum results is taken into account as a
systematic uncertainty. 

At the lightest pion mass $m_\pi = 285$ MeV we only have data for $\beta = 3.30, 3.57$ and $3.70$, i.e. the finest lattice
spacing corresponding to $\beta = 3.85$ was beyond reach. Unfortunately in this case it turned out that the $\beta = 3.30$
lattice spacing could not be used for the continuum extrapolation of the chiral condensate and strange quark number
susceptibility because the resulting fits had bad $\chi^2/dof$ values.
Hence at this lightest pion mass point only 2 lattice spacings are used, $\beta = 3.57$ and $\beta = 3.70$ which of
course leads to continuum fits where the number of data points equals the number of parameters. For this reason our
continuum results for the lightest pion is not fully under control and we call them continuum {\em estimates} only. As
we will see from the continuum extrapolation of the Polyakov loop cut-off effects are very small in this quantity and
all 3 lattice spacings can be used for $m_\pi = 285$ MeV. Hence the continuum result for the renormalized Polyakov loop
is fully under control.

In order to check the robustness of our results we have reanalyzed the two new data sets corresponding to $m_\pi = 440$
MeV and $285$ MeV using the strategy in \cite{Borsanyi:2012uq}. There the systematic uncertainty was quantified by
considering randomly chosen nodes for the spline interpolation as well as performing $O(a)$ and $O(a^2)$ fits to the
continuum simultaneously. Comparison of the two methods for all 3 cases is again reassuring and shows that the
statistical and systematic effects have been estimated correctly.

On all figures below the results from the second interpolation/extrapolation method is used. Note that even though
monotonous interpolations are used and the continuum extrapolated central values are also monotonous, the
errors on the central values are temperature dependent and may lead to a non-monotonous error band. This
does happen in some cases.

The renormalized light chiral condensate is shown in figure \ref{pbp} for all 3 pion masses and
the strange quark number susceptibility is shown in figure \ref{qsusc} again for all three pion masses while the
renormalized Polyakov loop is shown in figure \ref{ploop}. 
In each case the solid band shows the result of our continuum extrapolations. 

\begin{figure}
\begin{center}
\includegraphics[width=11cm]{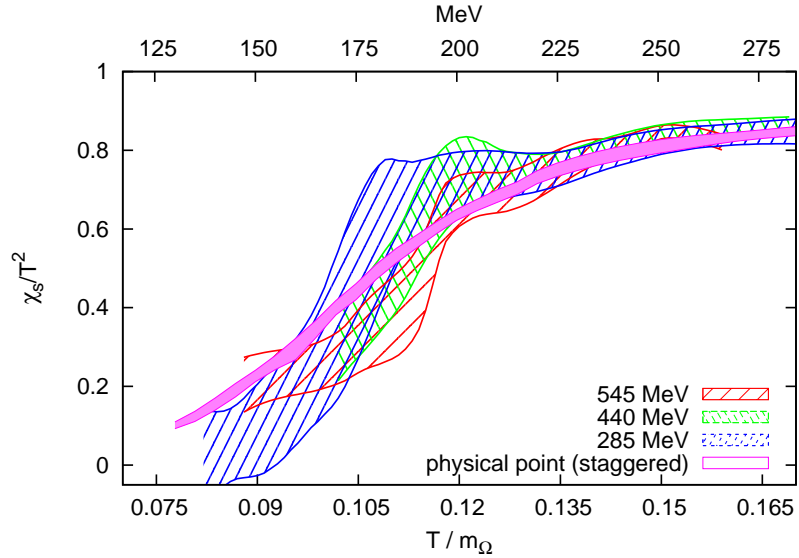}
\caption{Comparison of the continuum strange quark number susceptibility 
results for the three pion masses $m_\pi = 545$ MeV, $m_\pi = 440$ MeV and $m_\pi = 285$ MeV. The continuum result at the physical
point is also shown, obtained from staggered simulations.
\label{qsusc_cont}}
\end{center}
\end{figure}

\begin{figure}
\begin{center}
\includegraphics[width=11cm]{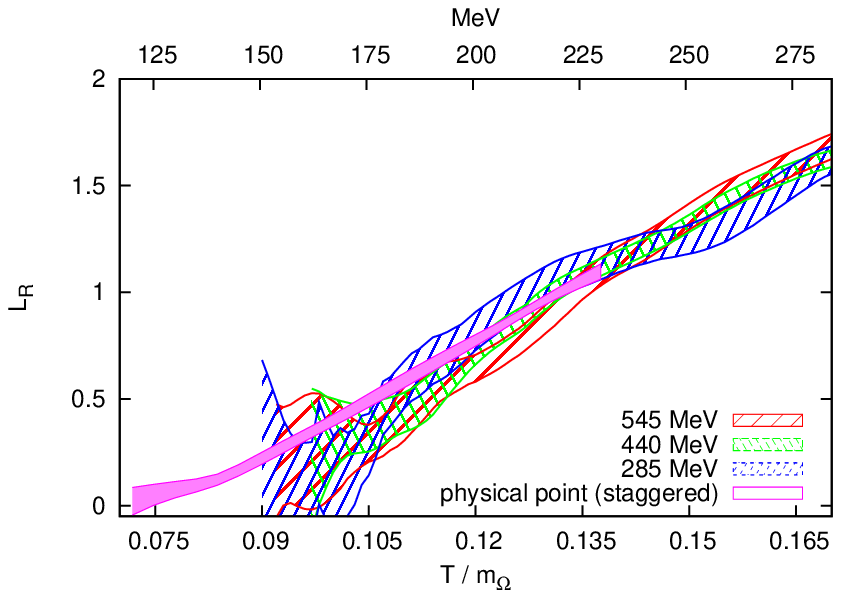}
\caption{Comparison of the continuum renormalized Polyakov loop 
results for the three pion masses $m_\pi = 545$ MeV, $m_\pi = 440$ MeV and $m_\pi = 285$ MeV. The continuum result at the physical
point is also shown, obtained from staggered simulations.
\label{ploop_cont}}
\end{center}
\end{figure}

Once continuum results are obtained at each of the three pion masses these continuum results can be compared for each
observable. Clearly, the pseudo-critical temperature defined by the chiral condensate is decreasing with decreasing pion
mass, see figure \ref{pbp_cont}. The pseudo-critical temperatures corresponding to the strange quark number
susceptibility and Polyakov loop on the other hand are only mildly sensitive, if at all, to the pion masses, see figure
\ref{qsusc_cont} and \ref{ploop_cont}. On these comparison plots we also show the result of past investigations using
the staggered formulation where continuum extrapolated results were possible to obtain at the physical pion mass 
\cite{Borsanyi:2010bp}.

Clearly, the staggered physical and continuum results fit nicely into the trend observed for the Wilson results: the
pseudo-critical temperature corresponding to the light chiral condensate is decreasing further with decreasing pion mass
while the strange quark number susceptibility shows only mild or no dependence. This is presumably because the strange
quark in the valence sector is the one dominating the strange quark number susceptibility and the light quarks enter
only through their sea contribution. On the other hand the light chiral condensate depends on the pion mass through both
the sea and valence sectors.

\section{Summary and outlook}
\label{summary}

In this paper we continued our program of lattice QCD thermodynamics using the Wilson fermion formulation. Our previously
published results at a relatively heavy pion mass $m_\pi = 545$ MeV was extended by including two ligther pions, $m_\pi
= 440$ MeV and $m_\pi = 285$ MeV. Our main goal
was to investigate the pion mass dependence of several observables which may be used to define a pseudo-critical
temperature. The continuum extrapolation was fully under control for $m_\pi = 440$ MeV but since we only used two
lattice spacings for $m_\pi = 285$ MeV, in the latter case we refer to it as a continuum estimate only (except for the
Polyakov loop where three lattice spacings were used and the result is hence fully under control). In any case the
continuum results support the picture that emerged from staggered simulations: the pseudo-critical temperature
associated with the light quarks is much more sensitive to the pion mass than the pseudo-critical temperature associated
with the strange quark. The light chiral condensate may be used to define the former and the strange quark number
susceptibility may be used to define the latter. The Polyakov loop which becomes an order parameter in the infinitely
massive quark limit also shows little pion mass dependence.

\begin{figure}
\begin{center}
\includegraphics[width=11cm]{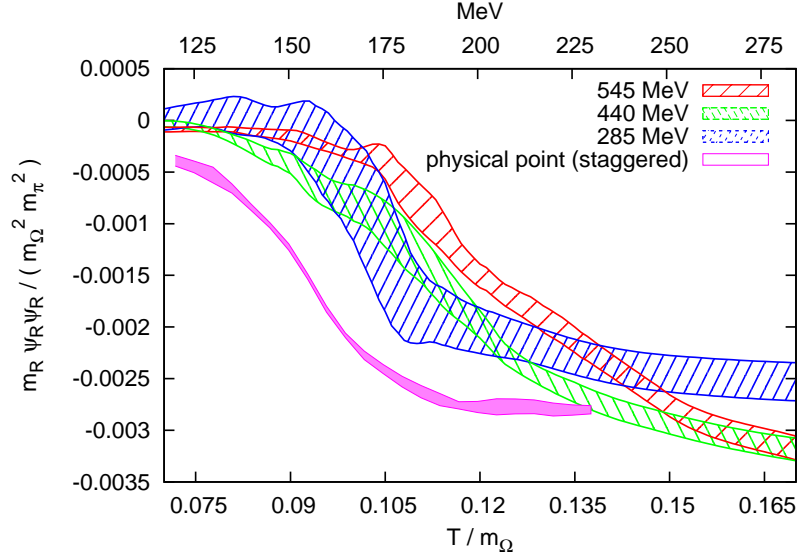}
\caption{Comparison of the continuum renormalized chiral condensate 
results for the three pion masses $m_\pi = 545$ MeV, $m_\pi = 440$ MeV and $m_\pi = 285$ MeV. The continuum result at the physical
point is also shown, obtained from staggered simulations.
A downward shift in the pseudo-critical temperature with decreasing pion masses is clearly visible.
\label{pbp_cont}}
\end{center}
\end{figure}

We see a clear decrease in $T_c$ obtained from the light chiral condensate as the pion mass decreases and not much
sensitivity to the pion mass in $T_c$ obtained from the strange quark. The physical pion mass is beyond reach for
our simulations with Wilson fermions however with staggered fermions these are readily available. The comparison of our
3 Wilson continuum results corresponding to $m_\pi = 545$, $440$ and $285$ MeV with the staggered continuum result at physical
pions confirm this picture further as the light chiral condensate curve as a function of temperature moves further to
the left as $m_\pi = 285$ MeV decreases to the physical point. However the strange quark number susceptibility as a
function of temperature is only mildly sensitive to the pion mass.

One may attempt to extrapolate the chiral condensate to the physical point but judging from figure \ref{pbp_cont} the
result will have a rather large uncertainty and therefore will not be very informative. In order to obtain results with
reasonable accuracy at the physical point a simulation at at least one lattice spacing directly at the physical point 
will be needed.

\section*{Acknowledgment}

Computations were carried out on both GPU clusters \cite{Egri:2006zm} at the
University of Wuppertal, Germany and Eotvos University, Budapest, Hungary and also on the BG/Q supercomputer in
Forschungszentrum Juelich, Germany.

This work was supported by the EU Framework Programme 7 grant 
(FP7/2007-2013)/ERC No 208740, by the Deutsche Forschungsgemeinschaft
grant SFB-TR 55 and by the grant OTKA-NF-104034 by OTKA.

\end{document}